\documentclass[twocolumn,prl,aps]{revtex4-1}
\usepackage{xspace,amsmath,amsfonts,amsthm,amssymb,amsbsy,graphics,color,graphicx,bbm,soul}
\usepackage[dvipsnames]{xcolor}
\usepackage[breaklinks=true]{hyperref}

\hypersetup{
  colorlinks   = true, 
  urlcolor     = blue, 
  linkcolor    = blue, 
  citecolor   =  red 
}

\newcommand{\bra}[1]{\left\langle{#1}\right|}
\newcommand{\ket}[1]{\left|{#1}\right\rangle}
\newcommand{\op}[2]{\ket{#1}\!\!\bra{#2}}
\newcommand{\expt}[1]{\left\langle{#1}\right\rangle}
\newcommand{\ip}[2]{\left\langle{#1}\right|\left.{#2}\right\rangle}

\newcommand\jced[1]{{\color{blue} \st{#1}}} 

\begin{document}
\title{How the result of a single coin toss can turn out to be 100 heads}
\author{Christopher Ferrie}
\author{Joshua Combes}

\affiliation{
Center for Quantum Information and Control,
University of New Mexico,
Albuquerque, New Mexico, 87131-0001}

\begin{abstract}
We show that the phenomenon of anomalous weak values is not limited to quantum theory.  In particular, we show that the same features occur in a simple model of a coin subject to a form of  classical backaction with pre- and post-selection.  This provides evidence that weak values are not inherently quantum, but rather a purely statistical feature of pre- and post-selection with disturbance. 
\end{abstract}

 \date{\today}
\maketitle

In many quantum mechanical experiments, we observe a dissonance between what actually happens and what ought to happen given na\"{\i}ve classical intuition.  For example, we would say that a particle cannot pass through a potential barrier---it is \emph{not allowed} classically. In a quantum mechanical experiment the ``particle''  can ``tunnel'' through a potential barrier---and a paradox is born. Most researchers spent the 20th century ignoring such paradoxes (that is, ``shutting up and calculating'' \cite{Mer08a}) while a smaller group tried to understand these paradoxes \cite{Eve56a,DeW08a,Boh52a,Bel64a} and put them to work \cite{DeuJoz92a}.

Experimentalists can probe the quantum world is through measuring the expectation value of an observable $A$. After many experimental trials the expected value is  
\begin{align}\label{def:expectation}
\expt{A}_\psi = \expt{\psi | A | \psi}
\end{align}
where $\ket{\psi}$ is the quantum state of the system under consideration.  The measurement of such an expected value allows us to demonstrate, for example, that Bell's inequalities \cite{Bel64a} are violated. Thus measurement of the expected value can have foundational significance.

In Eq.~(\ref{def:expectation}) the potential values one can observe are limited to the eigenvalue range of $A$.  It was surprising, then, that Aharonov, Albert, and Vaidman \cite{Aharonov1988How} claimed the opposite.  In 1988, they proposed the {\em weak value} of an observable. The weak value of $A$ is defined as \cite{Aharonov1988How,Duck1989Sense}
\begin{equation}\label{def:weak value}
a_w = \frac{\langle \phi | A | \psi\rangle}{\langle \phi | \psi\rangle},
\end{equation}
where $|\psi\rangle$ and $|\phi\rangle$ are called \emph{pre-} and \emph{post-}selected states.  Notice that when $\langle \phi | \psi\rangle$ is close to zero, $a_w$ can lie far outside the range of eigenvalues of $A$ hence the title of \cite{Aharonov1988How}: ``How the result of a measurement of a component of the spin of a spin-1/2 particle can turn out to be 100''.  When this is the case, the weak value is termed \emph{anomalous}.

Weak values are said to have both foundational and practical significance.  On one hand, they are claimed to solve quantum paradoxes \cite{Aharonov2005Quantum}, while on the other, they are claimed to amplify small signals to enhance quantum metrology \cite{Hosten2008Observation}  (but compare to \cite{Knee2013Quantum,TanYam13a,FerCom14a,KneGau14a,ComFerJia13a,ZhaDatWal13a}). One research program in the weak value community is to examine a paradoxical quantum effect or experiment and then calculate the weak value for that situation. Often the calculated weak value is anomalous. From this we are supposed to conclude the paradox is resolved (see, for example, \cite{Shikano2012Theory} for a recent review). So it would further seem, then, that anomalous weak values, if not \emph{the} source of quantum mysteries, provide deep insight into finding it. Indeed, since their inception, weak values have inspired deep thinking and debate about the interpretation and foundational significance of weak values \cite{Leg89a,Per89a,AhaVai89a,mermin1995limits,LeiSpe05a,Mer11a}.

Where a classical explanation exists no quantum explanation is required.  This is a guiding principle for quantum foundations research.  In this letter we provide a simple classical model which shows anomalous weak values are not limited to quantum theory.  In particular, we show the same phenomenon manifests in even the simplest classical system: a coin.  This shows that the effect is an artifact of toying with classical statistics and disturbance rather than a physically observable phenomenon.

Let us begin by defining the weak value as it was formally introduced before casting it into a more general picture.  We have a system with observable $A=\sum_a a |a\rangle\!\langle a|$ and meter system with conjugate variables $Q$ and $P$ so that $[Q,P]=i$.  The system and meter start in states $|\psi\rangle$ and $|\Phi\rangle=(2\pi \sigma^2)^{-1/4} \int dq' \exp(-q'^2/4\sigma^2)\ket{q'}$ and we define $\Phi(q)= \langle q |\Phi \rangle$. They interact via the Hamiltonian $H = A\otimes P$, then are measured in the bases $\{|\phi_k\rangle\}$ and $\{|q\rangle\}$ where $q\in(-\infty,\infty)$.  We are interested in the joint probability distribution of this measurement: $\Pr(q,\phi \,|\, \psi,\Phi) = |\langle \phi|\langle q |e^{-i x H/\hbar}|\psi\rangle|\Phi\rangle|^2$ where $x$ is the product of the coupling constant and interaction time.  In this case, it can be shown (as in, Chap. 16 of \cite{Aharonov2005Quantum}), in the limit $\sigma\to\infty$ \cite{Note0},
\begin{equation}
\langle \phi|\langle q |e^{-ix A\otimes P/\hbar}|\psi\rangle|\Phi\rangle = \langle \phi|\psi\rangle \Phi(q- x a_w),
\end{equation}
where $a_w$ is the weak value given in \eqref{def:weak value} and, assuming $a_w$ is real \cite{Note1},  is the average shift of the meter position given the states  $|\psi\rangle$ and $|\phi\rangle$.  Consider the following example.  We take the system observable $A=Z$, the Pauli $Z$ operator, and pre- and post-selected states
\begin{align}
\ket \psi & = \cos\theta/2 \ket{+1} +\sin\theta/2\ket{-1},\\
\ket \phi & = \cos\theta/2 \ket{+1} -\sin\theta/2\ket{-1},
\end{align}
where $\ket{+1}$ and $\ket{-1}$ are the $+1$ and $-1$ eigenstates of $Z$, respectively.
A short calculation reveals
\begin{equation}\label{eq:quant wv}
a_w =\frac1{\cos\theta},
\end{equation}
thus when $\theta \approx 1.5608$ we have $a_w=100$! This is patently non-classical as the states required to observe a value $a_w>1$, say, are in different bases.  
Next, we will show how to obtain an anomalous weak value from a system-meter picture and statistical conditioning. 

A large class of measurements that we can perform on a quantum system can be described by a set of Kraus operators and their corresponding positive operator valued measure (POVM). Below we will need to measure a coarse graining over a set of Kraus operators: a quantum operation.  We expand the unitary to first order in $x$: $U(x) = \exp(-ix A\otimes P  / \hbar)\approx \mathbb I\otimes  \mathbb I -ixA\otimes P  / \hbar$. To this order in perturbation theory, the Kraus operators for a position measurement on the meter are $M_q = \expt{q|U(x)|\Phi} =  \mathbb I  \ip{q}{\Phi}  -ixA \bra{q}P\ket{\Phi}{ /\hbar}$. Using $P= -i\hbar\, \partial /\partial q$ and $\partial_y\exp(-y^2/4\sigma^2) =( -y /2\sigma^2 )  \exp(-y^2/4\sigma^2)$ the Kraus operator becomes
\begin{align}
M_q = \left [ \mathbb I  -q   \frac{x }{2\sigma^2} A \right ]\Phi(q),
\end{align}
where $\sigma^2$ is the initial variance of the Gaussian meter state and $x$ is the coupling constant. Now we consider coarse grained measurements so that $q\leq0$ is identified as the ``$+1$'' outcome of $A$ and $q>0$ is identified as the result ``$-1$'', then the corresponding quantum operations are 
\begin{align}
\mathcal E_+ \rho= \int_{-\infty}^0\! dq\, M_q \rho M_q^\dag \quad {\rm and} \quad \mathcal E_- \rho = \int^{\infty}_0\! dq \, M_q \rho M_q^\dag.
\end{align}
Such quantum operations have conditional states $\rho_\pm = {\mathcal E_\pm \op{\psi}{\psi}}/{{\rm Tr}[\mathcal E_\pm \op{\psi}{\psi}]}$, which are generally mixed states.
Performing the integral gives the operation 
\begin{align}
\mathcal E_\pm \rho &=\frac{1}{2} \left [\rho  \pm  \frac{x }{\sqrt{2\pi\sigma^2}}\left (A \rho + \rho A\right ) \right ].
\end{align}
Collecting the constants we define $\lambda\equiv   2 x /\sqrt{2 \pi \sigma^2}$. Also we find it convenient to introduce a classical random variable, $s\in\{\pm1\}$ for the sign of the outcome. With these conventions the operation becomes
\begin{align}\label{eq10}
\mathcal E_s \rho &=\frac{1}{2} \left [\rho  +s  \frac{\lambda }{2}\left (A \rho + \rho A\right ) \right ].
\end{align}
Note that as $\lambda\to0$ the measurement approaches the trivial one, conveying no information and leaving the post-measurement unaffected.

The trace of Eq. (\ref{eq10}) describes the outcome statistics of weak measurement of the operator $A$ in the state $\rho$. This can be seen from the probability of observing the outcome $s$
\begin{align}\label{eq:quant weak}
\Pr(s|\psi) = {\rm Tr}\left [ \mathcal E_s \op{\psi}{\psi} \right ] = \frac 1 2 (1 +s \lambda \expt{\psi| A |\psi})
\end{align}
which is correlated with the expectation value of the operator $A$.   

Following Ref.~\cite{GarWisPop04} we now calculate the conditional expectation of the random variable $s$ given the pre- and post-selected states $\ket{\psi}$ and $\ket{\phi}$ respectively
\begin{align}
\!\!\mathbb E_{s|\phi,\psi} \left[ s \right]  & = \sum_{s=\pm1}  s \frac{\Pr(s,\phi|\psi)}{\Pr(\phi|\psi)} = \sum_{s=\pm1} s \frac{\langle \phi|\mathcal E_s (\op \psi \psi )|\phi\rangle}{|\langle\phi|\psi\rangle|^2},
\end{align}
 where $\mathbb E_{x|y}[f(x)]$ denotes the conditional expectation of $f(x)$ given $y$ and $\Pr(\phi|\psi) = \sum_s \Pr(s,\phi|\psi) ={ \sum_s} \langle \phi|\mathcal E_s (\op \psi \psi )|\phi\rangle$ becomes $|\langle\phi|\psi\rangle|^2$. Expanding the numerator we obtain
\begin{align}
\mathbb E_{s|\phi,\psi} \left[ s \right]  & = \sum_{s=\pm1} \frac{s}{2 } \frac{\langle \phi|  |\psi\rangle\!\langle \psi|  +(s\lambda/2)  \{ |\psi\rangle\!\langle \psi| ,A \}_+  |\phi\rangle}{|\langle\phi|\psi\rangle|^2},
\end{align}     
where $\{A,B\}_+=AB+BA$. This result can also be arrived at using Bayes rule to determine $\Pr(s|\phi,\psi)$, which is known as the ``ABL rule'' in quantum theory (after Aharonov, Bergmann, and Lebowitz \cite{AhaBerLeb64a}). Further expanding the numerator \cite{Note1} we arrive at
\begin{align}
\mathbb E_{s|\phi,\psi} \left[ s \right]  & = \sum_{s=\pm1} \frac{s}{2} \left(1+ s\lambda \frac{\langle \phi| A|\psi\rangle}{\langle\phi|\psi\rangle} \right) ,\\
&= \lambda \,\frac{\langle \phi| A|\psi\rangle}{\langle\phi|\psi\rangle}.\label{eq:lambda}
\end{align}
Thus the conditional expectation of $s$ results in a quantity proportional to the weak value. 
Since the constant of proportionality is $\lambda$, to arrive directly at the weak value we consider the conditional expectation of the random variable $s/\lambda$. Using Eq.~\eqref{eq:lambda}, we have
\begin{align}
\mathbb E_{s|\phi,\psi} \left[\frac s \lambda\right]  = \frac1{\lambda}\mathbb E_{s|\phi,\psi} \left[s\right] = \frac{\langle \phi| A|\psi\rangle}{\langle\phi|\psi\rangle}.
\end{align}
Thus, an equivalent definition of the weak value is
\begin{align}\label{eq:weak value class}
a_w =\mathbb E_{s|\phi,\psi} \left[\frac s \lambda\right].
\end{align}
To relate this to the meter picture note that $s/\lambda =  s\sqrt{2\pi \sigma^2}   /2x$. Thus the limit of $\lambda\rightarrow 0$ is identical to $\sigma \rightarrow \infty$ \cite{Note0}. It is clear from (\ref{eq:weak value class}) that a weak value is a calculated quantity, specifically it is the conditional expectation of the random variable
$s/\lambda$.

From Eq.~(\ref{eq:quant weak}) we can see that
\begin{align}
\langle \psi | A|\psi \rangle  = \mathbb E_{s|\psi}\left[\frac s \lambda\right]= \sum_s  \frac s \lambda \Pr(s|\psi). 
\end{align}
By the classical law of total expectation we have:
\begin{align}
\langle \psi | A|\psi \rangle & = \mathbb E_{s|\psi}\left[\frac s \lambda\right] = \mathbb E_{\phi|\psi}\left[\mathbb E_{s|\phi,\psi}\left[\frac s \lambda\right]\right].
\end{align} 
From Eq.~(\ref{eq:weak value class}) we know we can replace $\mathbb E_{s|\phi,\psi}\left[\frac s \lambda\right]$ with the weak value, thus
\begin{align}
\langle \psi | A|\psi \rangle =  \mathbb E_{\phi|\psi}\left[\frac{\langle \phi | A | \psi\rangle}{\langle \phi | \psi\rangle}\right] .
\end{align}
So, the weak value arises close to the way it is often envisioned to---as a condition expectation---but to define it properly, we need to include a renormalization by the weakness parameter $\lambda$.

Now we demonstrate that it is possible to find anomalous weak values for pre- and post-selected states in the same basis provided there is classical disturbance.  In particular, we take $A=Z$, $\ket\psi = \ket{+1}$  and later we will postselect on $\ket\phi = \ket{-1}$.  Using the probabilities in Eq.~\eqref{eq:quant weak}, the probability of the outcome of the weak measurement is
\begin{equation}\label{weak s}
\Pr(s|\psi{ = +1})= \frac12(1+s\lambda).
\end{equation}
Since the measurement is in the same basis as the state, the state is unchanged and the final weak value will not be anomalous.  Thus, we must do something more.  To simulate the disturbance, we now apply a bit-flip channel which conditionally depends on the strength and outcome of the weak measurement. This is reasonable as one would expect, from quantum measurement theory, that the amount of disturbance should depend on the measurement. After the channel, the state becomes
\begin{equation}
 \op{+1}{+1} \mapsto (1-p) \op{+1}{+1} + p \op{-1}{-1},
\end{equation}
{ where $p$ is the probability of a bit-flip error. To match the quantum case, we want $p$ to be close to 0 when the weak value ought to be large (as the occurrence of large weak values is rare) and close to 1 when the weak value ought to be small. Such a functional form of $p$ is as follows:
\begin{equation}
p = \frac{1+s\lambda - \delta}{1+s\lambda},
\end{equation}
where $\delta$ is the \emph{disturbance parameter} which is constrained to be $0<\delta<1-\lambda$ so that $0<p<1$.  The particular form of $p$ is not important---many choices will lead to anomalous weak values. { One can even choose the $p$ here so that it is identical to effective $p$ from the fully quantum calculation.  Here we have introduced a new parameter $\delta$ not because we must, but because we can.}  We have chosen this form so the final expression is as simple as possible.  

In explicit probabilistic notation, we have
\begin{align}
\Pr(\phi = +1| s, \psi = +1) &= \frac{\delta}{1+s\lambda} \,\,(\text{no flip}),\\
\Pr(\phi = -1| s, \psi = +1) &= \frac{1+s\lambda-\delta}{1+s\lambda} \,\,(\text{flip})\label{eq:flip}.
\end{align}
Using Bayes rule, we find
\begin{align}
\Pr(\phi = +1,s| \psi = +1) &= \frac{\delta}{2} ,\\
\Pr(\phi = -1,s|  \psi = +1) &= \frac12(1+s\lambda-\delta).
\end{align}
Marginalizing over $s$, we obtain
\begin{align}
\Pr(\phi=+1|\psi=+1) &= \delta,\\
\Pr(\phi=-1|\psi=+1) &= 1-\delta.
\end{align}

We now have all the ingredients to calculate the weak value as defined in Eq.~\eqref{eq:weak value class}. An interesting case is when preselect  on $\psi =+1$ and post select on $\phi=-1$:
\begin{align}
a_w & =\mathbb E_{s|\phi,\psi} \left[\frac s \lambda\right],\\
&=\sum_{s=\pm 1} \frac{s}{\lambda} \frac{\Pr(s,\phi|\psi)}{\Pr(\phi|\psi)},\\
& = \sum_{s=\pm 1}  \frac s{2\lambda } \left(\frac{1+s\lambda-\delta}{1-\delta}\right),\\
& = \frac{1}{1-\delta}.\label{eq:quant wv2}
\end{align}
With $0<\delta<1-\lambda$, $a_w$ can take on arbitrary values, just as the quantum mechanical weak value in Eq.~\eqref{eq:quant wv}.  This is made obvious if we make a simple change of variable $\delta = 1 -\cos\theta$. 
}
The expression in Eq.~\eqref{eq:quant wv2} also illustrates the following important point: any disturbance whatsoever can result in an anomalous weak value.  Thus, the effect is solely that of disturbance, post-selection and renormalization.

Since the state here remains in the $Z$ basis at all times, this calculation is essentially classical.  To make this point unequivocally clear, we now give an explicitly classical protocol to realize anomalous weak values.  Our example revolves around a coin where the outcome ``Heads'' is associated with the sign ``$+1$'' while ``Tails'' is associated with the sign ``$-1$''. This allows us to compare the analysis above for a quantum coin case (a qubit) and a classical coin. As before we abstract the sign into a random variable $s$.

An efficient strong measurement of a coin after a flip will result in an observer measuring and reporting outcome $s$ with probability $\Pr(\text{report}\,s|\text{prepare}\, s)=1$. A classical weak measurement of the sign of a coin $s\in \{\pm 1\}$ means the observer did not properly ascertain if the coin was heads or tails. Such a measurement might arise from an observer not having the time to properly examine the coin or if there was oil on their glasses. We model this by introducing a probability $\Pr(\text{report}\, s|\text{prepare}\, s)=1-\alpha$   and $\Pr(\text{report}\,\neg s|\text{prepare}\,s)=\alpha$. To make the connection with the weak measurement in quantum coin case, see Eq.~(\ref{eq:quant weak}), we take $\alpha = (1-\lambda)/2$ so that 
\begin{align}\label{eq:class weak}
\Pr(s|\psi) &= \frac12(1+\lambda s \psi).
\end{align}
For a coin that starts in heads $\psi = +1$, so $\Pr(s|\psi=+1) = \frac12(1+\lambda s)$, which is identical to Eq.~\eqref{weak s}. In this case, the physical meaning of $\lambda$ is clear---it is strength of the correlation between the result $s$ and the preparation $\psi$.  

\begin{figure}\centering
  \includegraphics[width=1\columnwidth]{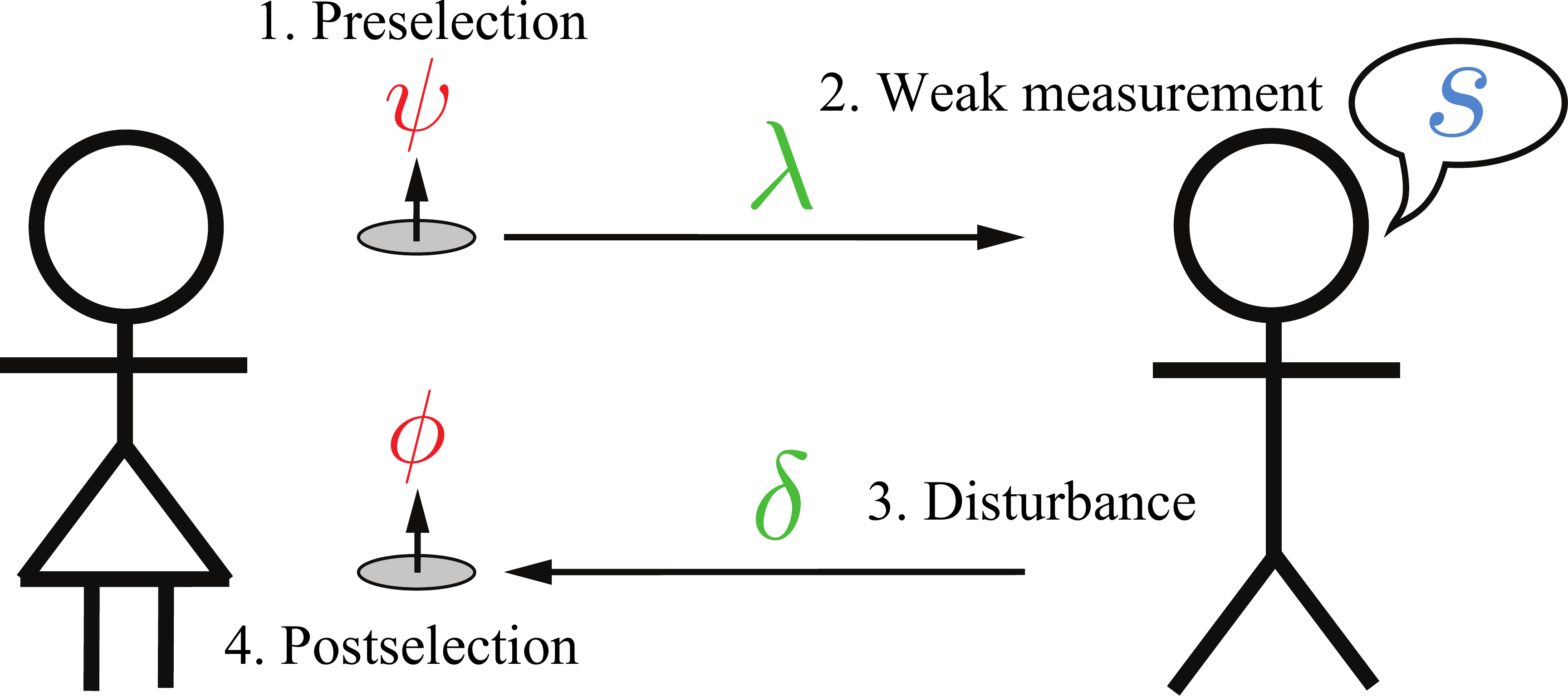}
  \caption{\label{fig:protocol} An illustration of the protocol used to realize anomalous classical weak values.}
\end{figure}

We now introduce a classical protocol directly analogous to the quantum protocol that produces anomalous weak values. There are two people, Alice and Bob.  The protocol is as follows (see also Fig.~\ref{fig:protocol}):
\begin{enumerate}
\item[1] {\bf Preselection}: Alice tosses the coin, the outcome $\psi$ is recorded, and she passes it to Bob.
\item[2] {\bf Weak measurement}: Bob reports $s$ with the probabilities given in Eq.~\eqref{eq:class weak}.
\item[3] {\bf Classical disturbance}: Bob flips the coin with probability given in Eq.~\eqref{eq:flip} and returns it to Alice.
\item[4] {\bf Postselection}: Alice looks at the coin and records the outcome $\phi$.
\end{enumerate}

For concreteness we preselect on heads, that is $\psi=+1$. Bob then makes a weak measurement of the state of the coin, which is described by Eq.~(\ref{eq:class weak}). In order to implement classical backaction we introduce a probabilistic disturbance parameter $\delta$ to our model. { Since we are free to choose this how we like, we choose the disturbance such that it results in the same flip probability in Eq.~\eqref{eq:flip}.

The point here is Bob will flip the coin (i.e. $+1\rightarrow -1$) with probability $1-\delta$. Although, $\delta$ can be thought of as a ``disturbance'' parameter,  a more entertaining interpretation is to think of Bob as an ``$\lambda$ liar, $\delta$ deceiver'':  Bob accepts the coin and lies about the outcome with probability $1/2(1-\lambda)$ and then further, to cover his tracks, flips the coin before returning it to Alice with probability depending on what he reports.

Since the probabilities for the classical and quantum case are identical, the weak value is identical:
\begin{align}
a_w = \frac1{1-\delta} \label{eq:class weak value}.
\end{align}
In particular, we see that the classical weak value can be arbitrarily large provided the parameter $\delta$ is close to $1$ and we pre-select $\psi=+1$ and post-select on $\phi=-1$.  Take the example $\delta = 0.99$.  The classical weak value of $s$, from Eq.~\eqref{eq:class weak value} with $\delta = 0.99$, is $a_w = 100$.  Thus, the outcome of the coin toss is 100 heads!    
}

Some remarks are in order.  First, we have pointed out that our model (in fact, any model) requires measurement disturbance for anomalous weak values to manifest.  Since, \emph{in theory}, classical measurements can have infinite resolution with no disturbance, some might consider our model non-classical.  However, \emph{in practice} classical measurements do have disturbance and do not have infinite precision.  While we have not provided a physical mechanism for the disturbance here, it is clear that many can be provided.  Thus, we leave the details of such a model open.  We note that in the context of Leggett-Garg inequalities, a similar observation was made: the weak value is bounded for non-invasive measurement \cite{WillJor08}.

The second, and perhaps more significant potential criticism, is that we have given a classical model where only \emph{real} weak values occur.  Whereas, the quantum weak value is a complex quantity in general.   It is often stated that weak values are ``measurable complex quantities'' which further allow one to ``directly'' access other complex quantities \cite{Dressel2013Understanding}.  However, the method to ``measure'' them is to perform separate measurements of the real and imaginary parts.  This illustrates that the weak value is actually a \emph{defined} quantity rather than a measured value\jced{d}.  Thus, we can easily introduce complex weak values in our classical model with two observable quantities and simply multiply one by the imaginary unit---not unlike descriptions of circular polarization in classical electromagnetic theory  (compare to the recent classical interpretation of a weak value experiment~\cite{BliBekKof13}).

In conclusion, our analysis above demonstrates a simple classical model which exhibits anomalous weak values.  Recall that the way in which weak values are used in foundational analyses of quantum theory is to show that they obtain anomalous values for ``paradoxical'' situations.  To suggest that this is meaningful or explanatory, it must be the case that such values cannot be obtained classically.  Here we have shown they can indeed.  Thus, the conclusion that weak values  can explain some paradoxical situation or verify its quantumness are called to question. Our results provide evidence that weak values are not inherently quantum, but rather a purely statistical feature of pre- and post-selection with disturbance. { Finally our work suggests an interesting question for future research, namely classical inference (including counterfactual reasoning) in the presence of classical disturbance.  }

{ 
{\em Remark}: after completion of this manuscript we were made aware of similar work on ``contextual values''~\cite{DreJor12}, where the authors reach quite a different conclusion from ours. 
}
\begin{acknowledgements}
The authors thank Carl Caves, Jonathan Gross, Asger Ipsen, Matt Leifer, Jacob Miller, Matt Pusey, Howard Wiseman and the Referees  for discussions about weak measurements and weak values. We also thank Justin Dressel and Andrew Jordan for making us aware of Refs.~\cite{WillJor08,DreJor12,BliBekKof13}.  This work was supported in part by National Science Foundation Grant Nos. PHY-1212445 and PHY-1314763 and by the Canadian Government through the NSERC PDF program.
\end{acknowledgements}

\end{document}